\documentclass[aps,prb,reprint,twocolumn,superscriptaddress,amsmath,amssymb,amsfonts]{revtex4-2}

\AtBeginDocument{\usepackage{booktabs}}               
\makeatletter
\g@addto@macro\bfseries{\boldmath}
\makeatother

\usepackage[varg]{txfonts}

\usepackage{color}
\usepackage{hyperref}
\definecolor{cL}{RGB}{32,145,140}
\hypersetup{colorlinks=true,linkcolor=cL,citecolor=cL,urlcolor=cL} 
\usepackage{multirow}
\usepackage{amsmath}
\usepackage{graphicx}
\usepackage[percent]{overpic}
\usepackage{mathrsfs}
\usepackage{bm}
\usepackage{braket}
\usepackage[nolist,nohyperlinks]{acronym}
\usepackage{natbib}
\usepackage{microtype}
\usepackage{float}
\usepackage[table,xcdraw]{xcolor}

\newacro{DM}[DM]{{Dzyaloshinskii-Moriya}}

\newcommand{\conbo}{CoNb$_2$O$_6$}
\newcommand{\mgconbo}{Mg$_{x}$Co$_{1-x}$Nb$_2$O$_6$}

\begin{document}
%
%
%
%
%
\title{Low Temperature Domain Wall Freezing and Non-Equilibrium Dynamics in the Transverse-Field Ising Model Material \conbo}

\author{C. L. Sarkis}
\affiliation{Department of Physics, Colorado State University, 200 W. Lake St., Fort Collins, CO 80523-1875, USA}
\author{S. S{\"a}ubert}
\affiliation{Department of Physics, Colorado State University, 200 W. Lake St., Fort Collins, CO 80523-1875, USA}
\author{V. Williams}
\affiliation{Department of Physics, Colorado State University, 200 W. Lake St., Fort Collins, CO 80523-1875, USA}
\author{E.S. Choi}
\affiliation{National High Magnetic Field Laboratory, Florida State University, Tallahassee, FL 32310-3706, USA.}
\author{T. R. Reeder}
\affiliation{Department of Physics and Astronomy and Institute for Quantum Matter, Johns Hopkins University, Baltimore, Maryland 21218, USA}
\author{H. S. Nair}
\affiliation{Department of Physics, University of Texas El Paso, 500 W University Ave, El Paso, Texas 79902}
\author{K. A. Ross}
\email{Kate.Ross@colostate.edu}
\affiliation{Department of Physics, Colorado State University, 200 W. Lake St., Fort Collins, CO 80523-1875, USA}
\affiliation{Quantum Materials Program, CIFAR, MaRS Centre,
West Tower 661 University Ave., Suite 505, Toronto, ON, M5G 1M1, Canada}

\date{\today}
%
%
%
%
\begin{abstract}
\conbo\ is a rare realization of the transverse field Ising model (TFIM), making it a useful tool for studying both equilibrium and non-equilibrium many-body quantum physics.  Despite a large body of work dedicated to characterizing this material, details of the ordered states in the presence of relatively weak transverse fields have not been discussed in detail. Here, we present a detailed study of \conbo\ via ac susceptibility measurements in order to further characterize its low temperature behavior in the presence of a transverse field. Specifically, we call attention to an unconventional freezing transition in zero-field below T$_F$ = 1.2~K, occurring \emph{within} the well-known commensurate antiferromagnetic (CAFM) state that onsets at $T_{N2}$ = 1.9~K.  We performed a series of transverse-field quenches into this frozen state, which resulted in a slowly relaxing susceptibility, $\chi^{\prime}(t)$, that followed a logarithmic decay within the time range measured.  We discuss the frozen state in the context of the freezing of previously discussed ``free'' chains arising from domain walls between the four degenerate sublattices of the CAFM state. We also attempted to observe Kibble-Zurek scaling by quenching the transverse field into the frozen state at different rates.  This produced a null result; the behavior can be fully explained by coarsening of domains over the timescale of the quenches.  The absence of a clear Kibble-Zurek scaling is itself surprising, given the proposed ubiquity of the phenomenon for general second order phase transitions.
\end{abstract}

\maketitle

\section{Introduction}
Recently non-equilibrium condensed matter has become a prominent field of study as a generator of exotic phenomena, such as many-body localization\cite{pal2010many,nandkishore2015many,PhysRevLett.114.170505} and Kibble-Zurek scaling\cite{zurek1985cosmological,dziarmaga2005dynamics,chandran2012kibble}. While theory has been propitious in some areas, much progress in non-equilibrium relies heavily on the guidance of experiment. Thus, experimental studies which are centered on tractable models are crucial for the field of non-equilibrium physics.

One of the archetypal models of condensed matter is the transverse field Ising model (TFIM)\cite{pfeuty1970one,elliott1970ising,dutta2015quantum},
\begin{equation}
    H = J\sum_{<i,j>}\sigma_{i}^{z}\sigma_{j}^{z} - h_{x}\sum_{i}\sigma_{i}^{x},
\end{equation}
where $\sigma_{i}^{\alpha}\ (\alpha=x,y,z)$ are Pauli spin matrices, $J$ is the exchange interaction strength, and $h_x$ represents an external magnetic field transverse to the Ising (\textit{z}) axis. This model can be exactly solved in 1D and features a quantum critical point at $J = 2h_x$\cite{pfeuty1970one,dutta2015quantum}. The TFIM provides a robust theoretical framework to study quantum criticality and non-equilibrium many-body physics. 

\begin{figure}[t]
    \centering
    \includegraphics[width=\columnwidth]{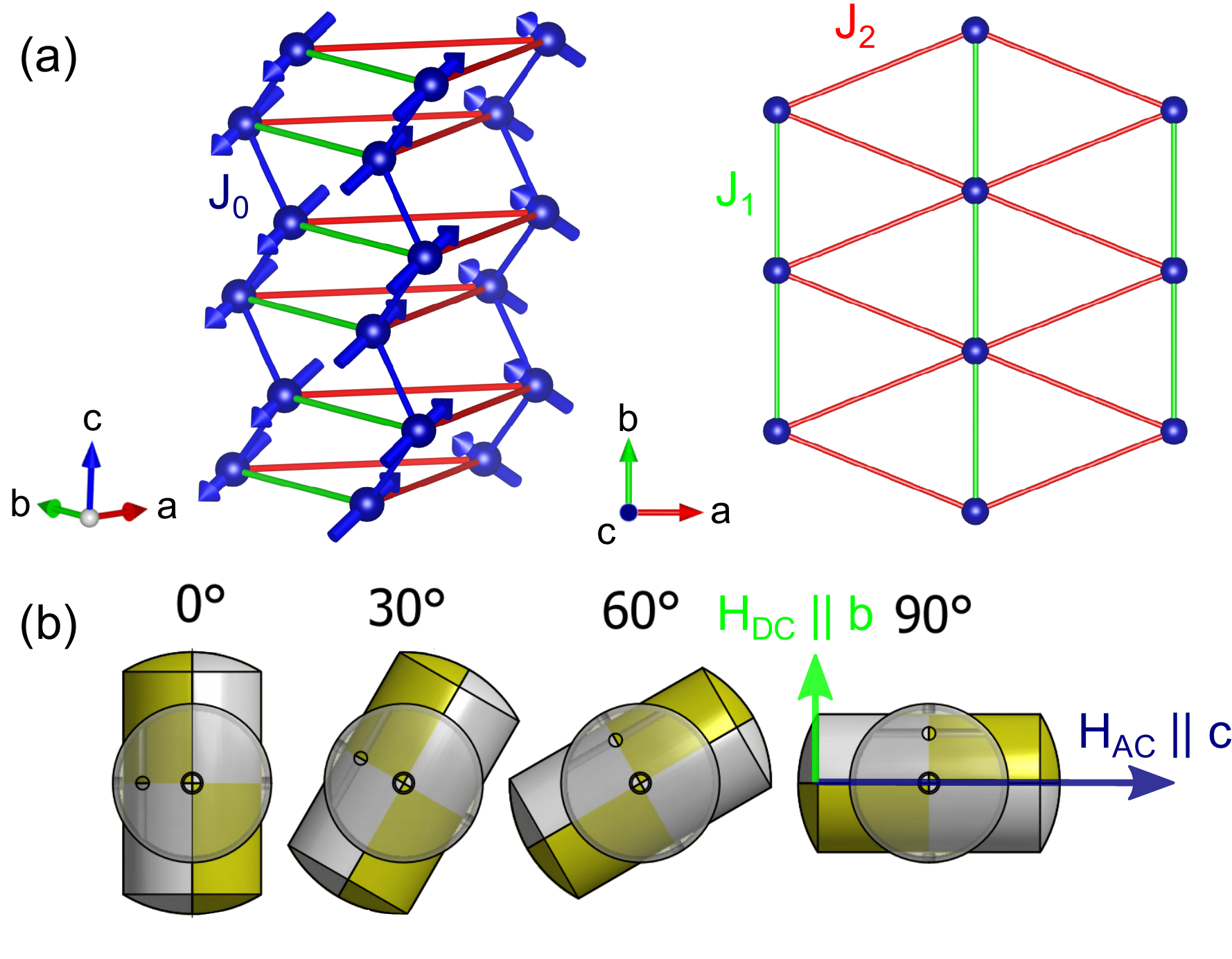}
    \caption{(a) Magnetic structure of \conbo\ in the CAFM state. Nearest neighbors form zig-zag chains along $c$ with ferromagnetic interactions ($J_0$). Second ($J_1$) and third ($J_2$) neighbors (both AFM) connect isosceles triangles in the $ab$-plane which stabilize 3D order at low temperatures.  $J_0$ is roughly an order of magnitude larger than the other exchange interactions, making \conbo\ quasi-1D. (b) Diagram of the ac-susceptometer at the NHMFL, showing the directions of the ac and dc-fields. The ac-field can be set at any angle with respect to the dc-field, within a single plane of rotation.  For our experiments, the ac field was oriented along the $c$-axis (the average Ising direction), and the dc field was usually oriented along the $b$-axis (transverse to the Ising axes).}
    \label{fig:structure}
\end{figure}

While real material analogs of the TFIM with critical fields accessible in current laboratory settings are rare, one key example is the quasi-1D Ising material \conbo. This material was extensively studied from the 1970's to 2000's, with the goal of characterizing its complex magnetic phase diagram in the presence of a magnetic field \cite{maartense1977field,scharf1979magnetic,hanawa1992disappearance,hanawa1994anisotropic,heid1995magnetic,kobayashi1995competition,heid1997magnetic}, though the \emph{transverse} field behavior was not investigated until later \cite{coldea2010quantum}.   The electronically insulating \conbo\ crystallizes into the orthorhombic Columbite space group  P$ b c n$, with room temperature lattice parameters of $a$ = 14.1475$~\AA$, $b$ = 5.712$~\AA$, and $c$ = 5.045$~\AA$ \cite{weitzel1976kristallstrukturverfeinerung}. The Co$^{2+}$ sublattice is shown in Fig. \ref{fig:structure}a. Co$^{2+}$ ions are linked by oxygen octahedra forming zig-zag chains along the $c$-axis, with $\sim$90$^\circ$ superexchange giving rise to ferromagnetic nearest neighbor interactions $J_{0}$ and a pronounced 1D behavior \cite{mitsuda1994magnetic}. Staggered planes of isosceles triangles form in the $ab$-plane, with antiferromagnetic (AFM) interactions $|J_{1}| > |J_{2}|$ between chains estimated to be an order of magnitude smaller than $J_{0}$\cite{kobayashi1999three}.   The isosceles triangular arrangement combined with AFM Ising interactions results in geometric frustration, which when combined with the 1D magnetism of the chains and strongly Ising-like interactions, leads to a complex magnetic phase diagram that strongly depends on the field direction \cite{lee2010interplay, da2007neutron,hanawa1994anisotropic}. In zero-field, the system undergoes a phase transition at $T_{N1}$ = 2.9~K into an incommensurate AFM (ICAFM) phase with temperature-dependent wavevector $q=(0~q_y~0)$. Below $T_{N2}$ = 1.9~K, the system locks into a commensurate AFM (CAFM) phase with $q = (0~0.5~0)$ via a first-order phase transition. Within the CAFM phase, neutron diffraction showed that \conbo\ has an ordered moment of $\mu \approx 3.2\mu_{\mathrm{B}}$\cite{heid1995magnetic}. Heat capacity measurements showed that magnetic contributions persist up to $T$~=~25~K \cite{hanawa1994anisotropic} and have an associated entropy of $R$ln2, supporting an $S_{\text{eff}} =\frac{1}{2}$ picture for the low temperature angular momentum degrees of freedom \cite{hanawa1994anisotropic,liang2015heat}.  In the CAFM state, there are two local moment directions, canted $\pm 31^\circ$ away from the $c$-axis and lying within the $ac$-plane. A series of papers by Kobayashi, \emph{et al.} reported anisotropic domain coarsening within the $ab$-plane (the triangular network) at $T$ = 1.5~K exhibiting an anomalously small growth exponent of $n$ = 0.2\cite{kobayashi1999anisotropic,kobayashi2004domain}.  This low growth exponent was attributed to the effects of frustration.

More recently, there has been renewed interest in \conbo\  due to its application to the TFIM\cite{lee2010interplay}. Neutron scattering under a transverse field (i.e. applied along the $b$-axis) revealed an emergent $\mathbb{E}_8$ symmetry\cite{coldea2010quantum}, which describes the 1D transverse field Ising chain at the quantum critical point with small but finite longitudinal field\cite{zamolodchikov1989integrals,rutkevich2010weak,kjall2011bound,lee2010interplay}. Transverse field heat capacity\cite{liang2015heat} and nuclear magnetic resonance \cite{kinross2014evolution} measurements revealed that the quantum phase transition (QPT) for an isolated Ising chain in \conbo\ would be at 5.25~T, however, this is buried within the 3D ordered phase which exhibits a QPT at 5.45~T\cite{da2007neutron}, due to weak but non-zero interchain couplings.  Recently it has been argued that additional off-diagonal exchange interaction terms are important for a quantitative understanding of \conbo \cite{fava2020glide,morris2020duality}, implying that the symmetry is not strictly Ising-like. Even so, the field-induced quantum critical point of the proposed models still maps onto the TFIM \cite{fava2020glide}. Given how well-studied the material now is due to its behavior in a transverse field, there is a surprising lack of detail in the literature about its response to low transverse field strengths. In order to use \conbo\ to experimentally investigate non-equilibrium properties of the TFIM, it is important to fully characterize its low temperature and low transverse field magnetic properties, particularly since they exhibit slow relaxation in this regime. 

Here, we report ac susceptibility measurements on \conbo\ taken down to a temperature of $\it{T}$ = 0.5~K.  We used a transverse dc field geometry (Fig. \ref{fig:structure} b) to measure the ac susceptibility along the $c$-axis in various scenarios: 1) the zero-field ac susceptibility, 2) the ac susceptibility as a function of transverse field, and 3) the relaxation of the ac susceptibility in zero-field after ramping the transverse magnetic field to zero at various rates ($r_Q$), spanning 0.1~T/min to 10~T/min.  For the zero-field  data we find good agreement with prior measurements \cite{hanawa1994anisotropic}, including the presence of a freezing transition which onsets at $T_F$ $\sim$ 1.2~K in our sample.  This freezing transition, which occurs within the commensurate ordered phase, is not usually discussed in studies of \conbo, but appears to be present in several samples \cite{hanawa1992disappearance,hanawa1994anisotropic,kobayashi1999three, liang2015heat}.  The transverse-field dependence of the ac susceptibility at 0.5~K reveals several features which can be identified as transitions between different ordered states, culminating in the 3D quantum critical point (QCP) at 5.45~T. Upon quenching the transverse field into the frozen state, we observe a relaxation in the real ($\chi^{\prime}(t)$) and imaginary ($\chi^{\prime\prime}(t)$) components of the ac susceptibility, which is best described by a logarithmic decay.  This contrasts with the power law relaxation of ac susceptibility (and neutron Bragg diffraction) that was found by Kobayashi, \emph{et al.} at $T$ = 1.5~K after longitudinal ($c$-axis) field quenches \cite{kobayashi1999anisotropic,kobayashi2004domain}, and points to the role of a disordered potential leading to modified coarsening behavior below $T_F$.  Varying the quench rate ($r_Q$) of the transverse field, we also find a logarithmic dependence of $\chi'$ on $r_Q$. Though the scaling could also be fit fairly well to a power law, which would suggest a connection to the Kibble-Zurek mechanism (KZM), we find that the dependence can be attributed entirely to a systematic effect resulting from the relaxing population of domain walls over the course of the quench time. This highlights the care necessary in showing KZM behavior for experimental systems in which coarsening occurs on a similar timescale to the quench time\cite{biroli2010kibble,biroli2015slow}.  It appears that KZM is not evident in our measurements, which in itself is surprising given the ubiquitous nature of the mechanism for second order transitions\cite{zurek1985cosmological,chandran2012kibble} and glass transitions\cite{liu2015universal,xu2017dynamic}.
\begin{figure*}
    \centering
    \includegraphics[width=\textwidth]{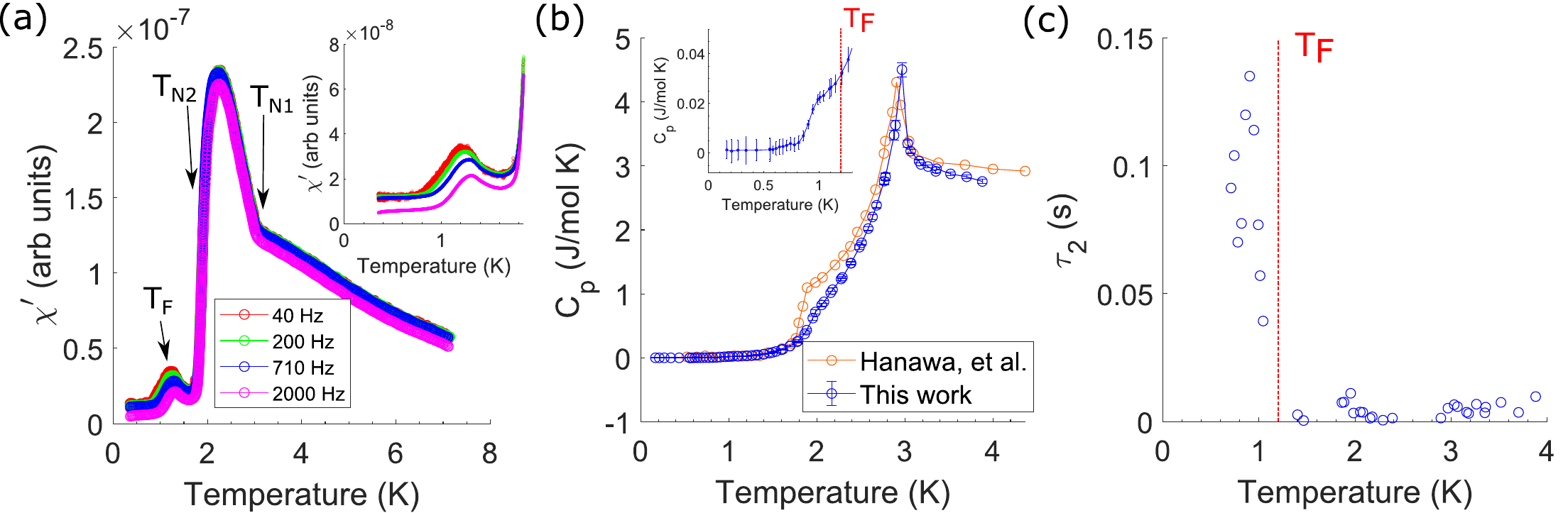}
    \caption{(a) Zero-field susceptibility of \conbo\ as a function of probe frequency. A frequency-dependent peak is seen centered around $T_F \sim$ 1.2~K (shown in inset), consistent with Ref. \onlinecite{hanawa1994anisotropic}. (b) Single crystal heat capacity of \conbo. A sharp peak at $T_{N1} = 2.9$~K coincides with the ICAFM transition, while a broad shoulder at $T_{N2} = 1.9$~K coincides with the CAFM transition. Heat capacity from Ref. \onlinecite{hanawa1994anisotropic} is also shown here in order to highlight the overall agreement between measurements. (inset) A small shoulder is observed in the heat capacity near $T = 1$~K, which could be due to the freezing transition observed in the zero-field susceptibility. (c)  Dependence of $\tau_2$ on temperature, which is a measure of the sample-to-stage relaxation processes. $\tau_2$ becomes non-zero near the onset of the frequency dependent low transition found in zero-field ac susceptibility.}
    \label{fig:data}
\end{figure*}


\section{Experimental Methods}
Small single crystals of \conbo\ (2 $\times$ 1 $\times$ 0.5~mm) were prepared via flux growth, following Ref. \onlinecite{wanklyn1976flux}, using 1.5~g \conbo with 1~g Borax (Na$_2$B$_4$O$_7$) in a Pt crucible (1250~$^{\circ}$C for 15~hrs then 750~$^{\circ}$C for 12~hrs with temperature ramps of 1.2~K/hr). Representative samples were crushed and checked for phase purity with a x-ray powder diffractometer. Alignment and crystallinity were checked with a Laue diffractometer, where the crystal was aligned within $\sim 0.5^\circ$. Two single crystals were used for ac susceptibility with masses of 13.6~mg and 8.2~mg, both cut into a cuboidal shape.

 Heat capacity measurements on a small single crystal (1.55~mg) grown from the same batch  were performed in zero field from 4~K to 0.06~K.  The data were taken in a Quantum Design Dynacool PPMS with dilution refrigerator insert using the thermal relaxation method.  

Transverse field ac susceptibility data were taken at the National High Magnetic Field Laboratory (NHMFL) in Tallahassee with a $^{3}$He cryostat mounted in a 20~MW 31~T 50~mm bore resistive solenoid dc magnet\cite{NHMFLsite}. The samples were subjected to both an ac and dc field, and were oriented such that the ac field was applied along the average moment direction ($c$-axis), while the dc field could be oriented anywhere within the $bc$-plane, i.e. between the average moment direction and the transverse direction (Fig \ref{fig:structure}b). The ability to change the angle \emph{in situ} enabled us to more accurately align the dc field to the $b$-axis within the plane of rotation, by observing the symmetry of the ac signal as a function of rotation angle (Appendix \ref{sec:TFS}).  The resistive magnet allowed for linear magnetic field ramps of the dc field with rates up to 10~T/min (note that a typical maximum ramp rate for a superconducting magnet is $\leq$ 1~T/min, which is what motivated us to use a resistive magnet instead). The ac field had amplitude 1.65~Oe and frequencies ranging between $f = 40$~Hz to 10~kHz, with most dc field-dependent measurements taken at 710~Hz. The dc field-dependent data were collected at $\it{T}$ = 0.5~K with high temperature background scans taken at $\it{T}$ = 10~K.

\section{Results \& Discussion}

\subsection{Temperature dependence of zero-field susceptibility}
The zero-field ac susceptibility data is shown in Fig. \ref{fig:data}a. A kink in the susceptibility at $T_{N1}$ = 2.9~K coincides with the ICAFM transition. An abrupt decrease in susceptibility is observed at the CAFM transition, $T_{N2}$ = 1.9~K. \conbo\ shows a pronounced frequency dependence below $T_{F}$ = 1.2~K accompanying a small peak in the susceptibility. This feature was also observed by Hanawa, \emph{et al.} in their ac susceptibility data, who attributed it to a metamagnetic (first order) transition and noted that it did not appear when measuring along the crystallographic $b$-axis\cite{hanawa1994anisotropic}.  The CAFM order is known to persist to 0.05~K \cite{da2007neutron}, so this frequency-dependent feature appears to signal a frozen state that co-exists with the CAFM state. 

The heat capacity of \conbo\ in zero-field is shown in Fig. \ref{fig:data}b), which agrees well with the literature\cite{hanawa1994anisotropic,liang2015heat}.  A sharp anomaly corresponds to $T_{N1}$, while a low temperature shoulder corresponds to $T_{N2}$. A second smaller shoulder is also observed around 1~K, which may be related to the freezing transition discussed above.  Significant magnetic heat capacity remains above the transitions and is known to persist to $\sim$25~K \cite{hanawa1994anisotropic}, well within the paramagnetic state, revealing the presence of short range spin correlations up to this temperature. The thermal relaxation method allows one to analyze the relaxation curves in terms of a two-time-constant model, where $\tau_1$ is related to the heat capacity of the sample, and $\tau_2$ is related to the heat flowing from within the sample to the sample platform and can be an indicator for slow relaxation within the sample\cite{lashley2003critical,suzuki2010accurate}.  The onset of significant  $\tau_{2}$ at $T\sim$ 1.2~K (shown in Fig. \ref{fig:data}c) is consistent with slow thermal relaxation within the sample, as expected based on the zero-field limit of the freezing transition revealed by ac susceptibility at the same temperature.

\begin{figure*}
    \centering
    \includegraphics[width=0.75\textwidth]{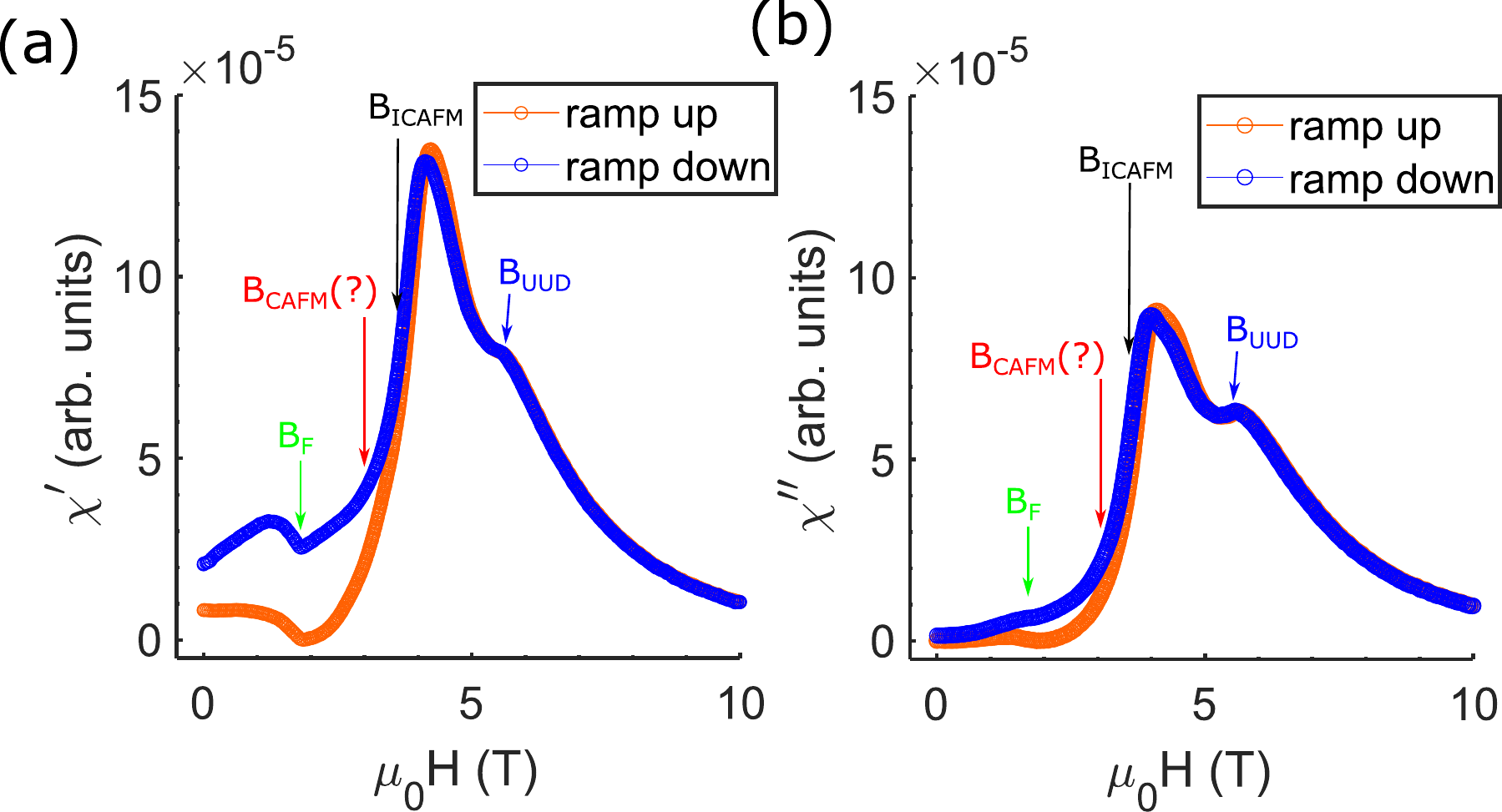}
    \caption{ Typical ac susceptibility data taken at $\it{T}$ = 0.5~K as a function of transverse field for real component (a) and imaginary component (b).  A shoulder observed near 5.3~T (B$_{UUD}$) in both real and imaginary components is consistent with the field induced transition between the paramagnetic and an ``up up down'' (UUD) ferrimagnetic phase identified in Ref. \onlinecite{da2007neutron}, which approximately corresponds to the TFIM QCP. A small hysteresis (history dependence) is evident at this phase boundary, but it is more pronounced below $\sim$3.7~T (B$_{ICAFM}$), which is a field-induced first-order transition into the ICAFM state \cite{da2007neutron}. The transition into the CAFM state reported in Ref. \onlinecite{da2007neutron} around 3~T (B$_{CAFM}$) is more difficult to observe in our susceptibility data, though a small kink in the decreasing real component can be observed. A ramp rate dependent discontinuity around $\sim$1.7~T (B$_F$) may represent the onset of the frozen phase.}
    \label{fig:phase_d}
\end{figure*}

Previous studies by Kobayashi, \emph{et al.} have also shown slow relaxation in \conbo\ within the CAFM phase, but at temperatures above the freezing transition that we report \cite{kobayashi1999anisotropic,kobayashi2004domain}.  In those studies, the FM $c$-axis chains were treated as single super-spins (since $J_0$ is by far the dominant exchange interaction, the spins along each chain were assumed to have the same orientation), and the defects of the ordering of those super-spins within the $ab$-plane was investigated. The CAFM state in \conbo\ has a four-fold degeneracy, so domains of this order populate randomly following a quench into the ordered state, with domain walls forming between them. Kobayashi, \emph{et al.} deduced the presence of ``free'' chains at certain types of domain walls, which arise due to the frustration of the isosceles triangular lattice formed by the chains. A time-dependence of both the correlation length and the magnetic susceptibility was also reported by Ref. \onlinecite{kobayashi2004domain} at $\it{T}$ = 1.5~K, just above the $T_{F}$. In that study, the relaxation behavior was modeled by a power law, consistent with domain coarsening theory.  However the growth exponent was found to be anomalously low ($n = 0.20\pm0.02$), in contrast to standard coarsening models of a four-fold degenerate Ising system with curvature-driven domain growth, which predict a growth exponent of $n$ = 0.5\cite{lifshitz1962kinetics,chowdhury1987domain}, though within the  anisotropic nearest-neighbor Ising (ANNNI) model used in Ref. \onlinecite{kobayashi2004domain} deviations from this exponent have been predicted\cite{cheon2001anisotropic}. 

One possible explanation for the freezing transition we observed within the CAFM state is the influence of disorder on the domain wall dynamics discussed above.  A natural route to disorder in \conbo\ comes from the CoO--Nb$_2$O$_5$ binary phase diagram \cite{burdese1964systems}, which shows that \conbo\ is not a ``line compound'', i.e. there is a fairly wide range of stoichiometries (within $\sim$ 1\% of the ideal 1:1 molar ratio) that lead to the same average crystal structure (the columbite structure type).  The relatively wide compositional space that stabilizes the columbite structure indicates a low enthalpy for defects relative to the concomitant gain in configurational entropy (e.g anti-site disorder or vacancies). Thus, even for perfect stoichiometry, one could expect a relatively high defect density \cite{maier1993defect} \footnote{Care must also be taken to convert cobalt oxide precursors to a single type. When purchased from chemical suppliers they often contain a mixture of Co$_3$O$_4$ and CoO which, if uncorrected, can lead to inaccurate Co stoichiometry.}. It is already known that \conbo\ can be extremely sensitive to small amounts of disorder \cite{sarvezuk2011suppression,nakajima2014magnetic}; for instance, in \mgconbo, less than 1\% substitution ($x$=0.008) is enough to entirely suppress the CAFM state at low temperatures\cite{nakajima2014magnetic}. 


\subsection{Transverse-field dependence of susceptibility}
Typical ac susceptibility data taken as a function of field at constant temperature (0.5~K) are shown in Fig \ref{fig:phase_d}.  The data were taken after cooling in zero-field from 2~K to 0.5~K. Features in these data align well with phase transitions identified from neutron scattering data \cite{da2007neutron} and are indicated in the figure by arrows. A shoulder near 5.3~T shows the transition between an ordered state and a field-polarized paramagnet, with a small accompanying hysteresis between ramps of increasing and decreasing field. The hysteresis is more dramatic below $\sim$ 3.7~T, where Ref. \onlinecite{da2007neutron} reported a transition between a ferrimagnetic phase and a field-induced ICAFM phase. A discontinuity in the real and imaginary components of the susceptibility appears near 1.7~T, with the field value at which it occurs weakly depending on quench rate, possibly signifying the boundary of the frozen state. 

Accompanying the development of hysteresis, a slow relaxation is observed in the susceptibility. The slow relaxation is seen for all fields below the ICAFM transition (3.7~T) but becomes more pronounced below the CAFM and freezing transitions ($< 3$~T) (see Appendix \ref{sec:aging}). In the frozen state, it appears that it would persist well beyond our maximum measurement time (600~s after reaching zero-field). 
\begin{figure*}
    \centering
    \includegraphics[width=0.75\textwidth]{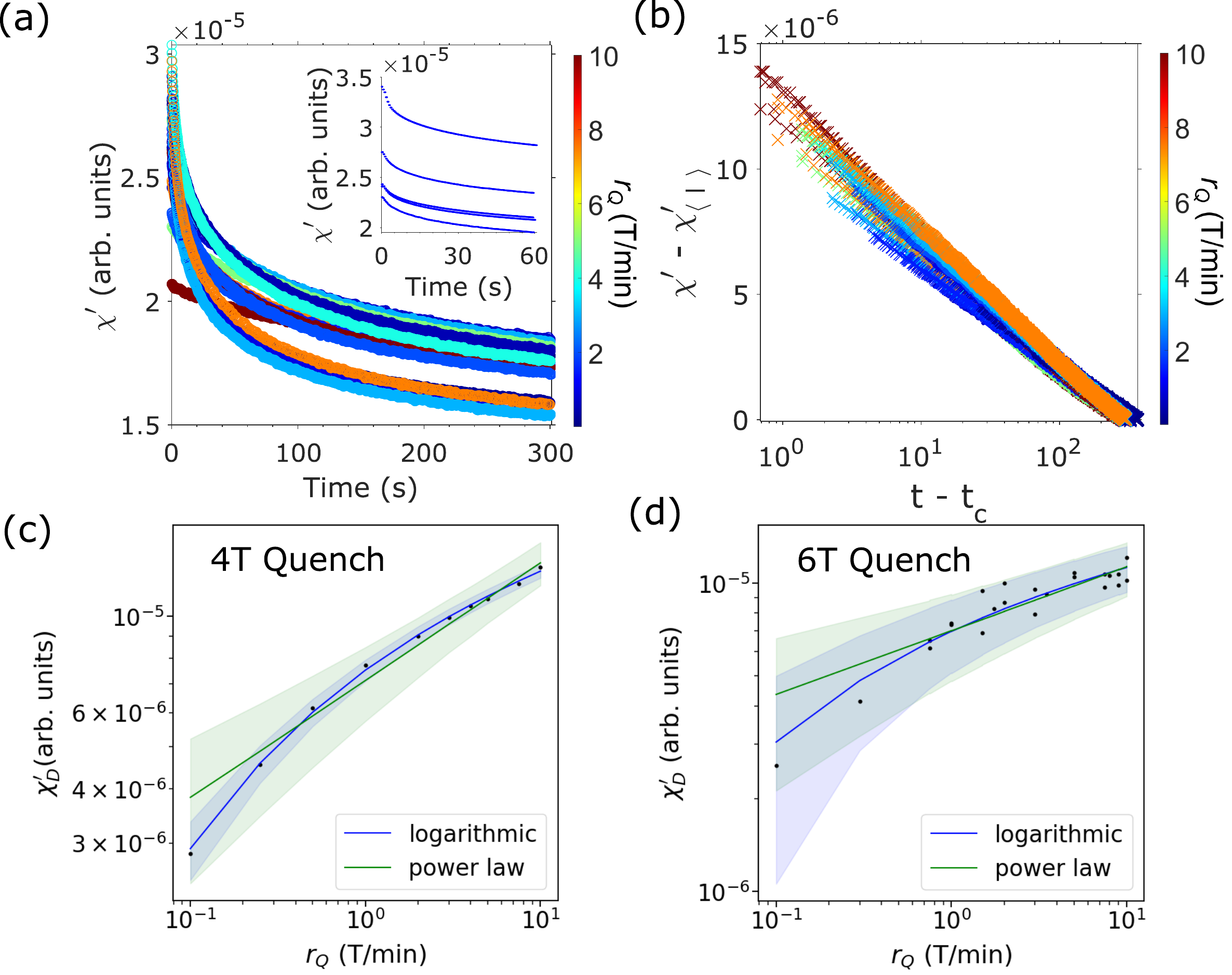}
    \caption{(a) Relaxation in zero-field at $\it{T}$ = 0.5~K after several quenches from a maximum field of 4~T at different rates, $r_Q$. Colors represent different $r_Q$, as quantified by the color bar. (inset) Zero-field relaxation of several quenches with $r_Q = 2$~T/min showing a history-dependence of the initial and final susceptibility, leading us to use the difference $ \chi^\prime_D $ as a measure of the relaxation at a given $r_Q$ (see Eqn. \ref{eqn:difference}). (b) Zero-field relaxation of 4~T quenches with a logarithmic time scale, where time is shifted by a critical time ``$t_c$'' defined as the time at which the field reaches 0.11~T. This produces straight lines, confirming the logarithm form of relaxation, but indicating an unusually low field for the onset of relaxation. (c) and (d)  $\chi^\prime_D $ vs $r_Q$ for a representative set of quenches (all completed after the same cool-down after warming past $T_F$) from a maximum field of 4T (c) and 6T (d). 
    The fits were performed using Bayesian non-linear regression.  The logarithmic form (Eqn. \ref{eq:systematic}) arises from purely systematic effects, while the power law form could potentially arise from the KZM. Shaded regions represent the 95\% credible intervals of the posterior predictive distribution.  For both sets of quenches, the logarithmic form gives the best fit.  }
    \label{fig:powerlaw}
\end{figure*}

In zero-field following a field quench, our time-dependent ac susceptibility is best fit to a logarithmic relaxation form:
\begin{equation}
        \chi^\prime (t) = a \ln{\bigg( \frac{t-t_0}{\tau}\bigg)} + c
        \label{eqn:log_relax}
\end{equation}
where $a$ is a negative scale factor, $t_{0}$ is the onset time of the relaxation, $\tau$ is needed to make the argument of the logarithm dimensionless and may represent an intrinsic relaxation timescale, and $c$ represents an offset that is partially due to a background contribution.  The form shown in Eqn. \ref{eqn:log_relax} produces an unbounded $\chi^\prime$ for infinite time and thus cannot represent the full relaxation curve.  We treat it as an early-time approximation for the relaxation, similar to the intermediate time logarithmic relaxation proposed by Ref. \onlinecite{brey2001slow}. Thus the $c$ parameter also likely accounts for some of the late-time behavior of the relaxation.  The functional form above provides a slightly better fit compared to a power law relaxation, and likewise, the power law fit gives a low exponent (-0.10 $\pm$ 0.03), hinting at the appropriateness of a logarithmic relaxation for our data (see Appendix \ref{sec:timedepanalysis}). 

The zero-field power law relaxation of ac susceptibility obtained in Ref. \onlinecite{kobayashi2004domain} was measured at 1.5~K, which is above $T_F$, and after different quench protocols compared to ours (they used temperature quenches and a $c$-axis field quench).  We therefore do not necessarily expect the relaxation forms to agree, particularly if the frozen state is due to disorder.  Indeed, disorder has been predicted to lead to logarithmic coarsening in Ising systems\cite{huse1985pinning}, while other studies have proposed a crossover from a power law to a logarithmic form beyond a critical time related to the length scale separating impurities\cite{chowdhury1987interface,chowdhury1987domain}. Slow relaxation in many systems, including spin glass systems, have also been modeled through hierarchically constrained dynamics \cite{mckay1982amorphously,hed2001spin,castellana2010hierarchical}. Such a dynamical structure has been shown to lead to either a stretched exponential\cite{palmer1984models,munoz1998hierarchical} or logarithmic\cite{brey2001slow} form of the relaxation.

\subsection{Quench-rate dependence of susceptibility}

One may think of the zero-field ac susceptibility as containing the response of the domain walls, as well as the intrinsic response of the CAFM ground state. Generally, we expect any signature which displays a relaxation over time to be partially attributed to the domain walls of the system (the units resulting from non-equilibrium).  However, one intriguing observation we have made is the presence of aging effects, i.e. certain properties of the relaxation depend on the overall history since cooling into the frozen state, \textit{even after} the transverse field is brought to 6~T, which is outside any ordered states.  This is discussed in more detail in Appendix \ref{sec:aging} (also see Fig. \ref{fig:intermediate}d and Fig. \ref{fig:params}).  While intriguing, this aging complicates the analysis of the quench rate dependence, since even identical repeated quenches lead to different offsets of the susceptibility, as shown in Fig. \ref{fig:powerlaw}a. To investigate the \emph{trends} of the relaxation within the frozen state, we thus characterized the relaxation by the difference between its initial value and final value
\begin{equation}
   \chi^{\prime}_{D} = \langle f \rangle - \langle l \rangle
    \label{eqn:difference}
\end{equation}
where $\langle f \rangle$ describes the average of the first second of zero-field data after the quench and $\langle l \rangle$ is the average of last second of collected data (usually at 300s). This represents the initial density of defects due to each new quench.   When $\chi^{\prime}_{D}$ is plotted as a function of quench rate ($r_Q$), a clear trend is seen (Fig. \ref{fig:powerlaw}c-d). In both crystals studied, this same type of dependence on quench rate was observed for all sets of quench data, which include quenches from 6~T across the QCP as well as quenches from 4~T which do not cross the QCP. Thus, the dependence on $r_Q$ is not due to the either the 1D or 3D QPT.  Rather, it seems to be due to a very low field transition around 0.11~T, which is made evident by the analysis described in Appendix \ref{sec:rqanalysis}. This field scale is also evident in the time-dependent data; when we shift the time axis by the time this field was reached ($t_c$), we see that $\chi^{\prime}(t)$ is linear on a log-log scale (Fig. \ref{fig:powerlaw}b). This also further emphasizes the appropriateness of the log relaxation form (Eqn. \ref{eqn:log_relax}) discussed above.

\begin{table}[]
\begin{tabular}{ll|l|l|}
\cline{3-4}
                                                                            &               & \multicolumn{1}{c|}{4~T} & \multicolumn{1}{c|}{6~T} \\ \hline
\multicolumn{1}{|l|}{\cellcolor[HTML]{DAE8FC}}                              & $a_L$ & 2.30(6)$\times 10^{-6}$                 & 1.9(2)$\times 10^{-6}$                  \\ \cline{2-4} 
\multicolumn{1}{|l|}{\multirow{-2}{*}{\cellcolor[HTML]{DAE8FC}Logarithmic}} & $B_c$         & 0.20(2)                 & 0.13(6)                 \\ \hline
\multicolumn{1}{|l|}{\cellcolor[HTML]{DEF6DE}}                              & $a_p$ & 7.1(3)$\times 10^{-6}$                   & 7.0(3)$\times 10^{-6}$                  \\ \cline{2-4} 
\multicolumn{1}{|l|}{\multirow{-2}{*}{\cellcolor[HTML]{DEF6DE}Power law}}   & $b$           & 0.27(2)                 & 0.21(3)                 \\ \hline
\end{tabular}
\caption{Fitted parameters for the quench-rate dependent data and models shown in Fig \ref{fig:powerlaw} c and d.  Bracketed numbers indicate one standard deviation.}
\label{tab:bayes}
\end{table}

Turning to the quench rate dependence of the difference in susceptibility, $\chi^{\prime}_{D} (r_Q)$ (Fig. \ref{fig:powerlaw}c and d), we used a Bayesian non-linear regression to fit two models; The first is a power law form, 
\begin{equation}
   \chi^{\prime}_{D}(r_Q) = a_p r_Q^b,
   \label{eq:KZM}
\end{equation}
where $a_p$ is a scale factor, and $b$ is the exponent.   This was motivated by the fact that an almost-linear dependence of $\chi^{\prime}_{D} (r_Q)$ appears on a log-log plot, and would potentially indicate KZM (see Appendix \ref{sec:kzm} for a discussion of the expected values for $b$ for this scenario). 

The second analytic form was derived considering a constant initial population of domain walls produced at a critical field $B_c$, which coarsen \textit{during} the quenches before the field reaches zero (see Appendix \ref{sec:rqanalysis} for more details).  This logarithmic form is, 
\begin{equation}
   \chi^{\prime}_{D}(r_Q) = a_L \ln\left[\frac{B_c r_Q}{t_m + B_c r_{Q}} \right],
   \label{eq:systematic}
\end{equation}
where $t_m$ is the measurement time between $\langle f \rangle$ and $\langle l \rangle$ (300 s for most quenches), $a_L$ is a scale factor that would be expected to be the same as $a$ as in Eqn \ref{eqn:log_relax}, and $B_c$ is a fitting parameter describing the critical field at which relaxation starts.

Both forms result in a reasonable fit of the data, but the logarithmic form, which we refer to as ``systematics'' (since it does not involve any additional physics such as KZM), does better at the lowest quench rates. The means of the fitting parameters and their standard deviations of the parameters for the two models are shown in Table \ref{tab:bayes}, and details of the fitting procedure and resulting joint distributions of parameters are in Appendix \ref{sec:rqanalysis}.  We note that the $\chi^{\prime}_D(r_Q)$ dependence of the 6~T quenches was significantly noisier than that of the 4~T quenches.  The reason for this is currently unknown, but it was a reproducible effect during the experiments.

 The power law model  would potentially indicate KZM, but with an unexpected scaling exponent of 0.27(2) for the 4~T data, and 0.21(3) for the 6~T data.  The naively expected exponents are greater than 0.5, but they can be reduced due to coupling to a bath (see Appendix \ref{sec:kzm}).  Yet, despite the predicted ubiquity for KZM across phase transitions, the logarithm explains our data better, as is evident from Fig \ref{fig:powerlaw}.  The form of the logarithmic scaling can be completely understood to be a consequence of a systematic effect, namely the coarsening of domains during the quench.  Another way to see this is that the fitting parameter $a$ in Eqn. \ref{eqn:log_relax}, which could reasonably be expected to scale with defect density, is independent of $r_Q$ (see Appendix \ref{sec:aging}).  
 
 Thus, it appears that we have not observed Kibble Zurek scaling in this experiment, which is somewhat surprising, since it is expected to apply at all second order phase transitions as well as glass transitions.  One clue that may help to explain this is the low critical field ($B_c < 0.3$ T) inferred from the systematics analysis (Eqn. \ref{eq:systematic}), which is well below any of the known phase transitions for a $b$-axis field applied to \conbo \cite{da2007neutron}.  Further, the time associated with crossing this field is also the offset in time required to produce linear dependence of the susceptibility on a logarithmic time scale (Fig. \ref{fig:powerlaw}b), i.e. $t_c$. 
 
 One possibility that could help to explain the low $B_c$, which also seems to explain the absence of Kibble Zurek scaling, is that we are primarily observing the effects of a transition that is due to a small component of the field being within the $ac$-plane, due to the slight misorientation of the sample (estimated to be about 1$^{\circ}$).  Indeed, \conbo\ is known to exhibit a complex phase diagram when the field is applied in the $ac$-plane, with several low field transitions are first-order in nature \cite{hanawa1994anisotropic}. Thus, it may be that the dominant behavior observed in our quench-rate experiments is due to coarsening effects that occur after crossing a \textit{first-order} phase transition induced by a small component of the field being applied in the $ac$-plane, rather than a second-order or glass transition.

\section{Conclusion}
 
In summary, we have performed ac susceptibility measurements of the quasi-1D Ising material \conbo\ in zero-field and under predominantly transverse magnetic fields. At low temperature (0.5~K) we see evidence of several transverse-field-induced transitions at fields below the known QCP (at 5.3~T), some of which are similar to those which have been predicted theoretically \cite{lee2010interplay} and observed in a prior study \cite{da2007neutron}.  We have also observed and characterized a zero-field transition into an unconventional frozen state at T$_{f}$ = 1.2~K, which is within the known commensurate antiferromagnetic state.  This is consistent with earlier reports of extremely slow dynamics in this temperature range, but has not been discussed in detail before. Within this frozen state, the $c$-axis ac susceptibility shows a logarithmic relaxation over time in response to a transverse ($b$-axis) dc field quench.  This form of relaxation is expected for coarsening of domains in the presence of disordered potentials, and this is an appealing explanation given prior work which successfully attributed some higher temperature behavior of \conbo\ to domain wall motion and coarsening \cite{kobayashi1999anisotropic,kobayashi2004domain}.  However, the presence of lattice disorder in our (and other group's) samples of \conbo\ remains to be confirmed.

We investigated the effect on the ac susceptibility response of ``quenching'' a transverse field across the various field-induced phase transitions of \conbo.  We found a surprising aging effect that indicates that the system is \textit{not} reset by applied a field of 6~T, which is above the QPT, within the ``quantum paramagnetic'' regime.   After accounting for this aging effect, we do find a distinct dependence of $\chi^\prime$ on quench rate, similar to a power law scaling that would be expected based on the KZM.  However, our observed dependence does not strongly depend on whether the QPT is crossed or not, and furthermore, we find it can be better fit by a logarithmic function arising from systematic effect, namely coarsening \textit{during} the field quench.  Thus, we observe no evidence for the KZM in quenches across the QPT, or across any of the field-induced transitions including the freezing transition.  Rather, the systematic analysis suggests that the observed effect is due to crossing a first-order transition brought about by a slight field misorientation.  

This ``null result'' is somewhat surprising given the proposed ubiquity of the KZM phenomenon for second order transitions (and generalized to glass transitions), and the fact that we clearly observe non-equilibrium states generated by the quenches. However, we cannot rule out that the Kibble Zurek scaling is being obscured by the aging effect we have observed.

Overall, our results emphasize the complexity of the dynamical behavior of the well-studied quasi-1D transverse field Ising model material, \conbo\, under relatively weak transverse fields.  This weak transverse field regime has until now not been explored in detail experimentally, but appears to contain a wealth of intriguing phenomena related to the frustration of the isosceles triangular lattice, which is likely to be strongly influenced by additional quantum fluctuations produced by the transverse field.

\begin{acknowledgements}
This research was funded by Dept of Energy grant DE-SC0018972. Authors would like to acknowledge Prof. Tarun Grover for insightful conversations relating to the Kibble-Zurek mechanism and domain coarsening effects. The authors would also like to thank Prof. James Neilson for stimulating conversations about defect chemistry in crystalline materials. A portion of this work was performed at the National High Magnetic Field Laboratory, which is supported by the National Science Foundation Cooperative Agreement No. DMR-1644779 and the state of Florida. Some figures were made using the 3D crystal modeling software VESTA\cite{momma2008vesta}.
\end{acknowledgements}
\appendix
\counterwithin{figure}{section}

\section{Kibble-Zurek considerations in \conbo}
\label{sec:kzm}
One goal of this study was to investigate the possibility of a Kibble-Zurek Mechanism (KZM) in \conbo. By quenching a control parameter through a continuous phase transition the system will inherently be driven out of equilibrium, a direct result of the divergence of the relaxation time near the critical point. This results in a non-zero density of defects which accumulate in the system for any finite quench rate. The density of defects, and all observables that depend on it, are predicted to scale as a power law with the quench rate, $r_Q$ \cite{chandran2012kibble}.  The density of defects, $\rho$, is predicted to scale as,
\begin{equation}
    \rho \sim r_{Q}^{d\nu/(1+z\nu)}
\end{equation}
where $r_{Q}$ is the quench rate, $d$ is the dimension, $\nu$ is the correlation length critical exponent, and $z$ is the dynamical critical exponent. The interest in Kibble-Zurek scaling is two-fold. First, despite being an inherently non-equilibrium effect, which are typically challenging to describe, this phenomenon is described \emph{entirely} by equilibrium properties (critical exponents). Second, the only requirement for Kibble-Zurek effects to occur is to cross a continuous phase transition at a finite rate, with the scaling relying only on the universality class of the system. While it was originally formulated by Kibble for defects in the early universe\cite{kibble1976topology,kibble1980some}, it was extended to condensed matter systems by Zurek, who proposed it in superfluid helium\cite{zurek1985cosmological}. It has since been generalized to quantum phase transitions\cite{dziarmaga2005dynamics,zurek2005dynamics,polkovnikov2005universal}, and experimental evidence for the KZM has been found in many systems including superfluid He\cite{hendry1994generation}, cold ion chains \cite{ulm2013observation}, and Bose condensates\cite{lamporesi2013spontaneous,anquez2016quantum}. To this date, no scaling from a KZM due to a quantum phase transition has been observed in a magnetic system, where a magnetic field can provide a natural tuning parameter. Due to strong spin-lattice coupling, typical relaxations in magnetic systems are fast (picoseconds). However, the timescale for observation of defects is not limited by this spin-lattice relaxation; rather, it is limited by the coarsening time, which relies on the mobility of defects (to some extent set by the exchange interactions) and the dimension.   Thus, one should expect that when defect motion is restricted, perhaps by disorder or frustration, KZ scaling could be readily observed over experimentally achievable timescales.  To accurately assess the power law exponent, observation of the scaling over several orders of magnitude of the quench rate is desirable. Conventional superconducting magnets  are limited to slow quench rates ( $<$ 10$^{-1}$ T/min ) do not offer a reasonable range.  Alternatively, resistive magnets can reach much higher magnetic field quench rates; the 31~T magnet at NHMFL provided us a range between 0.1~T/min and 10~T/min. Ultra-fast field ramps can also be achieved through pulsed magnets, which can reach quench rates of $>$ 10$^{3}$~T/min\cite{campbell1996nhmfl,NHMFLsite_pulsed}. 

The combination of an appropriate range of magnetic field quench rates as well as the observed slow coarsening dynamics of \conbo\ suggests it would have been a good candidate for observation of KZM. What is less obvious is what type of scaling should observed in \conbo. It is well-established that near the QCP \conbo\ displays hallmark behavior of the 1D TFIM with small but non-zero longitudinal field from the intrachain couplings\cite{coldea2010quantum,rutkevich2010weak,kjall2011bound}. In the case of defects being kinks along the Ising chains, scaling could be that of the QCP of the 1D Ising chain with an exponent of 0.5\cite{dutta2015quantum,mukherjee2007quenching}. However, \conbo\ orders into a 3D magnetically ordered ground state with much of its coarsening behavior at $T$ = 1.5~K attributed to domain walls within the $ab$-plane \cite{kobayashi1999anisotropic,kobayashi2004domain}. These defects would then perhaps be expected to scale according to the QCP of the 3D Ising model with an exponent of 0.75\cite{zheng1998determination,blote1980critical}. With a frozen state developing below $T_F$ = 1.2~K, the system may be described better by a 3D random field/bond Ising model, where dramatically increased dynamical exponents can greatly reduce expected scaling exponents. 

These predictions also are only inherently true for closed quantum systems, which we know \conbo\ is \emph{not}. Recent theoretical work for 1D Ising systems show that coupling to a bosonic bath greatly reduces the critical exponent down to 0.28\cite{oshiyama2020kibble}.  Experimental work on quantum simulators of the 1D TFIM have also estimated the value of the scaling exponent due to the KZM range between 0.20 and 0.33\cite{bando2020probing}. Coarsening and the KZM have been further explored\cite{biroli2015slow}, with estimates that in $d$ $>$ 2, coarsening does not strongly affect the expected critical exponents\cite{biroli2010kibble}.

Finally, we note that the scaling of the initial ac susceptibility with quench rate, parameterized by $a$ in Equation \ref{eqn:log_relax}, would likely be a good indicator of defect density.  But as discussed in Appendix \ref{sec:timedepanalysis} (and shown in Fig. \ref{fig:params}), we find that $a$ does not depend on quench rate systematically; rather, it appears to display some aging effects, though not as clearly as the $c$ parameter.  

\begin{figure}[t]
    \centering
    \includegraphics[width=0.85\columnwidth]{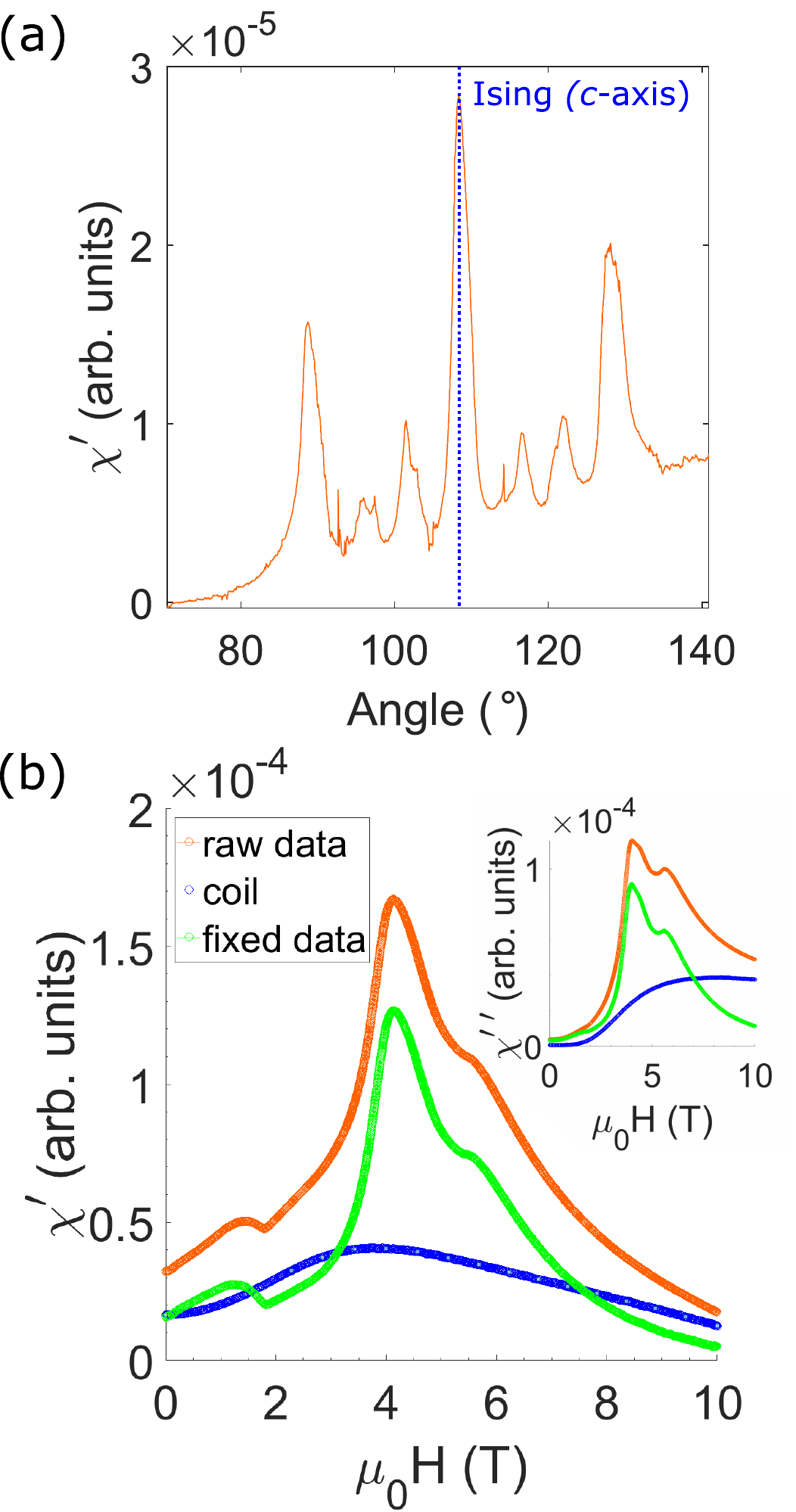}
    \caption{(a) Real component of ac susceptibility vs linear motor rotation angle taken under a 1~T dc field. A large peak near 108$^\circ$ shows the location of the average Ising axis ($c$), which is identified based on the symmetric peaks around it, which likely relate to field-induced phase transitions that are highly dependent on the field direction. (b) Example of transverse-field dependent susceptibility, with the background from coil shown, which was taken at 10~K for the real component (main figure) and imaginary component (inset) of the susceptibility. The background from the coil shows a frequency dependence as well as transverse field dependence but little to no quench-rate dependence at 710~Hz (not shown).}
    \label{fig:rotation}
\end{figure}

\section{Transverse Field Susceptibility}
\label{sec:TFS}
\subsection{Fine-tuning of orientation}
\label{sec:finetune}
For conventional ac susceptibility, one measures with the ac field in the same direction as the dc field. In contrast, for the susceptometer we used, the ac component was measured along the crystallographic $c$-axis, while the dc field could be applied anywhere in the $bc$-plane by an \emph{in situ} rotation of the ac coilset. The samples were loaded and aligned into the coilset with use of a Laue diffractometer, limiting a possible rotation of the sample out of the $bc$-plane to 1$^\circ$. In order to accurately align the dc field to be transverse to the Ising axis (i.e. $b$), we first located the $c$-axis by rotating the coilset under a 1~T dc field through a range of angles for which the direction of $H_{ac}$ is approximately equal to that of $H_{dc}$, and once determined, set the coilset to 90$^{\circ}$ from there.   Figure \ref{fig:rotation}a shows an approximately symmetric signal around ``108$^{\circ}$'',  which identifies the coil angle of 108$^\circ$ as the $c$-axis direction.  The several peaks observed around this position are likely indicating different field-induced transitions as a function of angle (for fields near the Ising axis, the phase diagram is known to be highly complex: \cite{hanawa1994anisotropic,kobayashi1999three}).  The overall background from the coil is not symmetric, but this is likely due to details of the coil design and centering in the magnet. 

\begin{figure*}
    \centering
    \includegraphics[width=\textwidth]{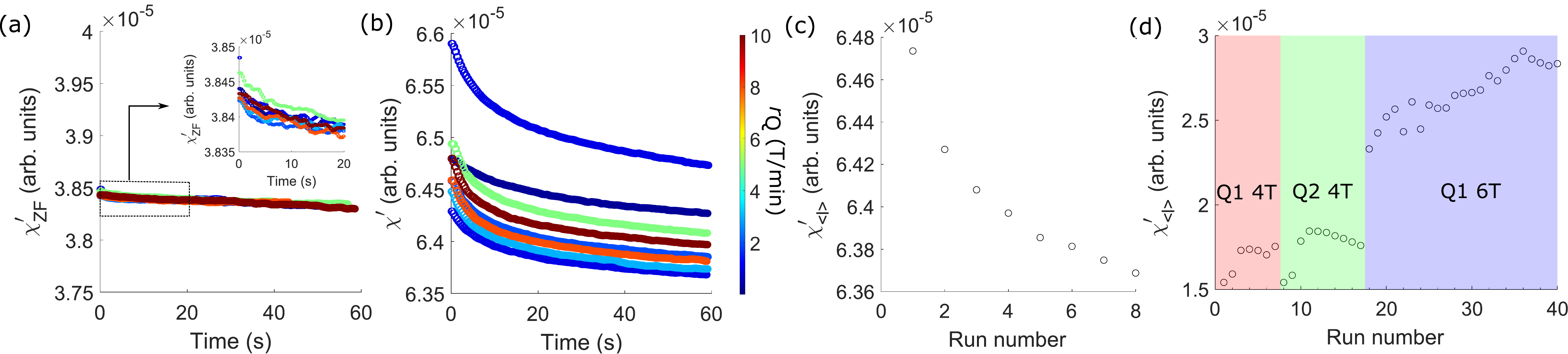}
    \caption{(a-b) Representative ac susceptibility data taken at the end of intermediate (stepwise) quenches for various points in the phase diagram. Color scale represents $r_Q$. (a) 4~T to 3~T quenches showing a small relaxation barely larger than instrument noise limit, showing the approach to the CAFM state near 3~T. (b) 3~T to 2~T quenches show a clear relaxation well beyond noise limit that exhibit a decrease in signal with increasing run number.  (c) Average susceptibility over last second of measuring time ($\chi^\prime_{<l>}$) as function of quench run number for data in panel b. Above the frozen transition, the system relaxes overall with each subsequent quench. (d) Average susceptibility over last second of measuring time (($\chi^\prime_{<l>}$)) as function of quench data for zero-field data showing aging effects within frozen state. Q1 4T and Q2 4T show two separate sets of quenches with a maximum field of 4~T, while Q1 6T shows a set of quenches following a maximum field of 6~T. }
    \label{fig:intermediate}
\end{figure*}

\subsection{Background subtraction}
\label{sec:background}
In general, for ac susceptometers in a transverse field, an anisotropic background signal of the coil can develop, which can be both field-dependent and frequency-dependent\cite{rucker2019compact}. For our setup, an estimate of the coil background was taken using a high temperature scan at 10~K,  well-into the paramagnetic phase (Fig. \ref{fig:rotation}b). Interestingly, we note that even at the maximum field of 10~T the imaginary component is still non-zero. This perhaps indicates that the system still exhibits frustration deep into the field polarized paramagnetic phase and could help to explain the aging effects we have observed (see next section). This background is subtracted from all ac susceptibility figures shown.


\subsection{Aging effects}
\label{sec:aging}
In an attempt to separate any possible trends coming from aging effects compared to quench rate effects, we instituted the following protocol for our magnetic field quenches: after the system was reset (warmed to 1.8~K then cooled in zero-field), we ran a series of quenches from 0~T up to a maximum field (either 6~T or 4~T) and then back down to 0~T. Each of the quench protocols consisted of an identical field sweep rate of 1~T/min from zero up to the maximum field, a hold at the maximum field for 10~s, and a field sweep (quench) rate chosen to be between 0.1~T/min and 10~T/min for the quench back to 0~T. The system was then reset again before another set of quenches by warming above $T_F$. By doing the quenches ``out of order'' with respect to quench rate, we are able to separate any overall aging effects (i.e. effects that correlate with run number, which loosely represents time since cooling into the frozen state) from effects depending on the quench rate, $r_Q$.

 To observe at what field the relaxation begins, we measured the susceptibility over time for many smaller intermediate magnetic field quenches (e.g. 4~T to 3~T, 3~T to 2~T etc.). A small relaxation on the order of the noise limit of our experiment is observed in the 4~T to 3~T quenches, shown in Fig. \ref{fig:intermediate}a. The onset of appreciable relaxation occurs for quenches from 3~T to 2~T, across the previously-identified transition between the ICAFM and CAFM ordered states at $\sim$ 3~T (Fig. \ref{fig:intermediate}b). For the final fields above the freezing transition, relaxations show an overall aging effect such that progressive quenches are shifted down with run number (Fig. \ref{fig:intermediate}c). For final fields below the anomaly at B$_f$ $\sim 1.7~T$, which we have tentatively associated with a field boundary of the frozen phase, the relaxation gains a more complicated dependence on the quench rate, shown in the main paper by the zero-field relaxation (Fig. \ref{fig:powerlaw}a). Looking at the average susceptibility at the last second of measurement time, shown in Fig. \ref{fig:intermediate}d, we notice two different aging behaviors in the 4~T and 6~T quenches. For the 4~T quenches, the average final susceptibility increases with initial runs and then saturates, while the 6~T quenches increase with consecutive quenches. Any trend that appears as a function of run number (which was not correlated with quench rate, by design) strongly suggests aging effects are present, and we find that these are surprisingly \emph{not erased} by going to 6~T (in the paramagnetic regime).  This may be related to the presence of significant energy dissipation up to 10~T, as indicated by the high $\chi^{\prime\prime}$ (Fig. \ref{fig:phase_d}) at 6~T, and suggests that one needs to go to higher fields to reset the system.

In order to account for the aging effect, which mainly seemed to produce an offset to $\chi^\prime$, we looked at the difference of the first second and last second of the susceptibility, $\chi^\prime(t)$ (Eqn. \ref{eqn:difference}).  

\subsection{Analysis of time-dependence of $\chi^\prime$}
\label{sec:timedepanalysis}
 
 After a magnetic field quench (i.e. when the dc field returns to zero), we observe a decay over time of the real and imaginary components of the susceptibility at $T$ = 0.5~K, i.e. a non-equilibrium state is generated. This decay appears for all quenches to zero field, regardless of their starting field values (which were either 4~T or 6~T), indicating that the non-equilibrium state is not due to quenching through the QPT associated with the 1D TFIM (at 5.2~T) or the 3D QPT to the field-polarized paramagnet (at 5.4~T).
 
 \begin{figure}[t]
    \centering
    \includegraphics[width=\columnwidth]{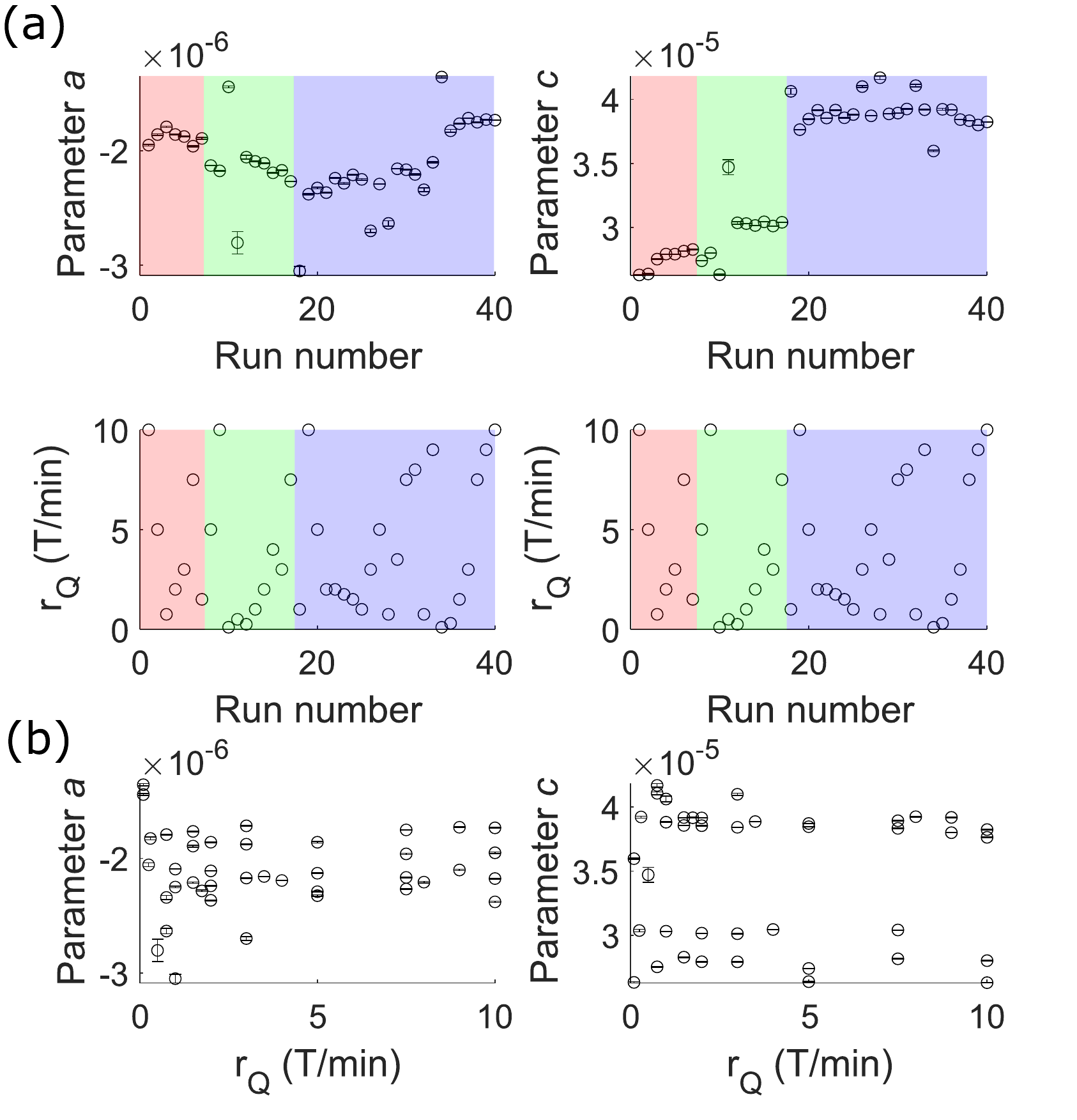}
    \caption{(a) Best fit parameters $a$ and $c$ for least squares fit of Eqn. \ref{eq:logrelaxappendix} as a function of quench run number, shown with psuedo-random $r_Q$ protocol as a function of quench run number. Color overlays shown distinct sets of quenches: red is Q1 4~T, green is Q2 4~T, and blue is Q1 6~T. Parameters $a$ and $c$ both show weak trends in run number, suggesting aging.  Parameter $c$ also shows a general offset to the data for each individual quench set (each set started after warming past $T_f$ and cooling back down in zero field).  (b) Plots of best fit parameters $a$ and $c$ for least-squares fit of Eqn. \ref{eq:logrelaxappendix} as a function of $r_Q$, where errors plotted are 95\% confidence intervals of the fitted parameters. Although the spread in values is much larger than the errorbars, they show no obvious dependence on ramp rate.}
    \label{fig:params}
\end{figure}
 
 Following a field quench at 0.5~K, the time-dependence of the zero-field relaxation within the frozen state can be fit relatively well to three different forms: a stretched exponential, a power law decay, and a logarithmic function given by,
\begin{subequations}
\begin{equation}
    \chi^\prime (t,B=0) = a \cdot \exp \left[-\left( \frac{t-t_0}{\tau} \right) ^\beta \right] + c
    \label{eq:stretchedexp}
\end{equation}
\begin{equation}
    \chi^\prime (t,B=0) = a \cdot \bigg(\frac{t-t_0}{\tau}\bigg)^n + c
    \label{power}
\end{equation}
\begin{equation}
    \chi^\prime (t,B=0) = a \cdot \ln\bigg(\frac{t-t_0}{\tau}\bigg) + c
    \label{eq:logrelaxappendix}
\end{equation}

\end{subequations}
\label{relaxationforms}
respectively. In each of the forms, $t_{0}$ represents the time of relaxation onset, $\tau$ represents an intrinsic time (needed in theory to make the time arguments dimensionless, but absorbed into the other parameters for fitting purposes), $a$ is a scale factor that one could expect to represent the initial population of defects generated by the quench, and $c$ is an offset due to late-time behavior and/or background. For the stretched exponential, $\tau$ represents the average relaxation time and $\beta$ represents the distribution of relaxation times. This form of relaxation is commonly found in glass systems\cite{mauro2018prony,potuzak2011topological}, and has also been found for hierarchically constrained dynamics\cite{palmer1984models}. A power law would instead indicate relaxations occurring on all time scales with growth exponent $n$ describing the coarsening.  Power law relaxation has been observed for \conbo\ at higher temperature above the glass transition where the exponent $n$ was found to be -0.2\cite{kobayashi1999anisotropic,kobayashi2004domain,nakajima2014magnetic}. Meanwhile, the logarithmic relaxation function has also been proposed for hierarchical dynamics\cite{brey2001slow}, and is often seen for domain coarsening in disordered models, such as the random field Ising model\cite{fisher2001nonequilibrium,corberi2015coarsening}.

\begin{figure}
    \centering
    \includegraphics[width=\columnwidth]{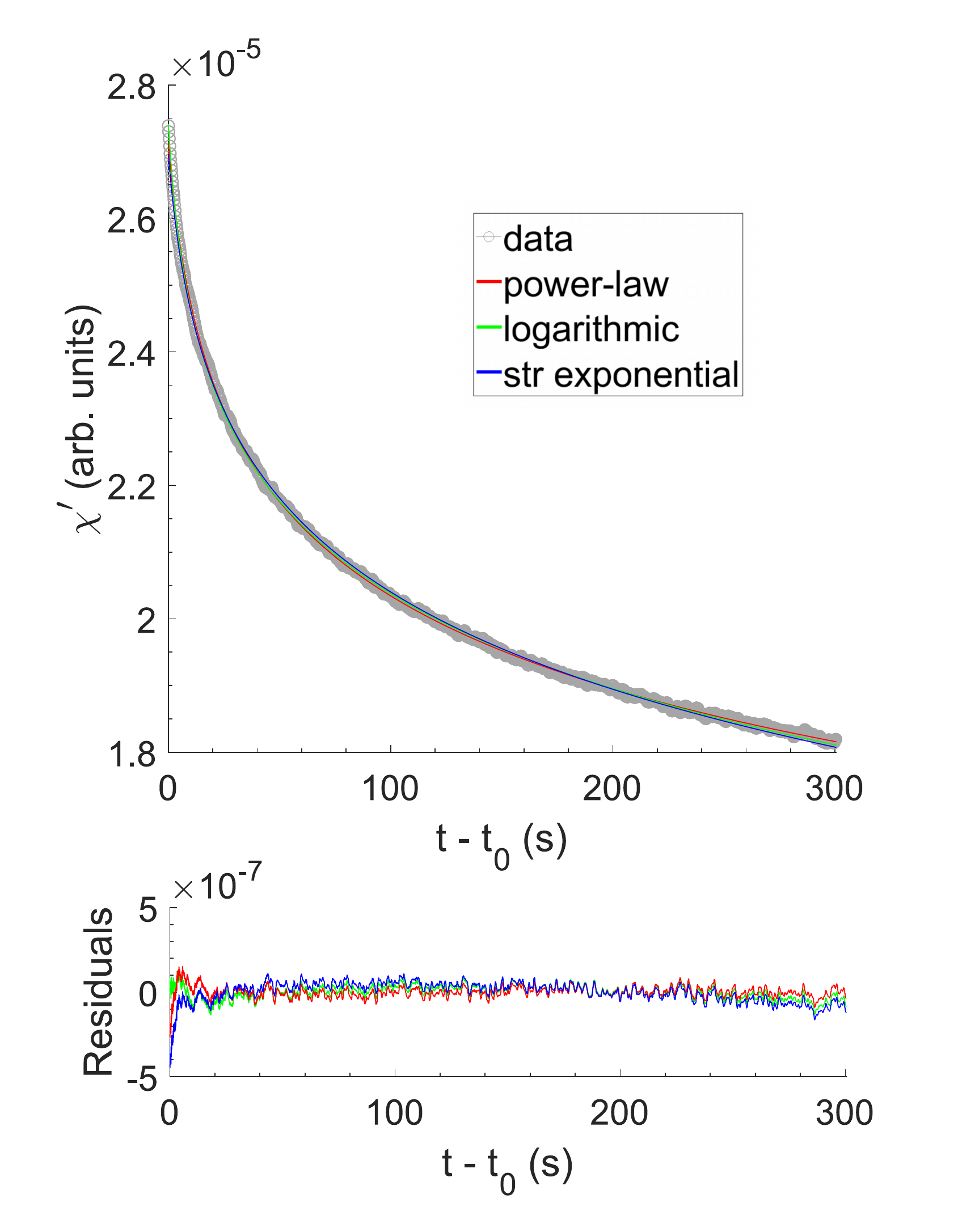}
    \caption{Example of least squares best fits to zero-field relaxation for the following functional forms: Power law, Stretched exponential and Logarithmic decay.}
    \label{fig:relax_form}
\end{figure}

To determine the best model for the decay, we compared $\chi^2$ for fits of the three models to several relaxation curves.  An example of best fits for the three functional forms are shown in Fig. \ref{fig:relax_form}, where the best fit parameters used are shown in Table \ref{tab:lsq_param}. Note that the form of the power law and logarithmic decay functions have absorbed $\tau$ into the definition of the other parameters, $a$ and $c$, as appropriate.  For the sake of comparison to Ref. \onlinecite{kobayashi1999anisotropic,kobayashi2004domain}, we note the data can be well-represented by a power law, but with a smaller exponent of $n = -0.097(3)$ compared to the value reported in literature ($n=-0.2$). While the $\chi^2$ for the three forms were comparable, in the end we chose the logarithmic form due to the reduced number of free parameters required to accurately fit the decay, and the very small exponent indicated by the power law fit.

\begin{table}[]
\begin{tabular}{c|c|c|cl}
\cline{2-4}
         & \multirow{2}{*}{\begin{tabular}[c]{@{}c@{}}Logarithmic\\ $a \cdot ln(t-t_0) + c$\end{tabular}} & \multirow{2}{*}{\begin{tabular}[c]{@{}c@{}}Power law\\ $a \cdot (t-t_0)^n +c$\end{tabular}} & \multicolumn{1}{c|}{\multirow{2}{*}{\begin{tabular}[c]{@{}c@{}}Stretched Exponential\\ $a \cdot exp \left[ -(\frac{t-t_0}{\tau})^\beta \right] + c$\end{tabular}}} & \multicolumn{1}{c}{} \\
         &                                                                                                &                                                                                             & \multicolumn{1}{c|}{}                                                                                                                                              & \multicolumn{1}{c}{} \\ \cline{1-4}
$a$      & -2.11(2)e-6                                                                                    & 3.54(6)e-5                                                                                  & 3.9(5)e-4                                                                                                                                                          &                      \\ \cline{1-4}
$t_0$    & 368.6(2)                                                                                       & 365.3(4)                                                                                    & 367(1)                                                                                                                                                             & \multicolumn{1}{c}{} \\ \cline{1-4}
$c$      & 3.02(1)e-5                                                                                     & -2.2(8)e-6                                                                                  & -1.2(3)e-4                                                                                                                                                         & \multicolumn{1}{c}{} \\ \cline{1-4}
$n$      & -                                                                                              & -0.097(3)                                                                                   & -                                                                                                                                                                  &                      \\ \cline{1-4}
$\tau$   & -                                                                                              & -                                                                                           & 124(8)                                                                                                                                                             &                      \\ \cline{1-4}
$\beta$  & -                                                                                              & -                                                                                           & 0.01(5)                                                                                                                                                            &                      \\ \cline{1-4}
$\chi^2$ & 3.2e-6                                                                                       & 2.3e-6                                                                                    & 7.0e-6                                                                                                                                                           &                      \\ \cline{1-4}
\end{tabular}

\caption{Table of least squares fits to typical zero-field relaxation of the susceptibility of \conbo\ at $T$ = 0.5~K taken after a 2~T/min quench from a peak field of 4~T. Best fit parameters are shown for three different forms of the relaxation: a logarithmic relaxation, a power law relaxation, and a stretched exponential relaxation. $\tau$ was absorbed into $a$ and $c$ for the power law and logarithmic forms respectively in order to reduce free parameters.}
\label{tab:lsq_param}
\end{table}


From intermediate quench data (quenches ending at a non-zero field), the system only shows a sizeable decaying population below the CAFM transition, B$_{CAFM}$ $\sim$ 3~T. Within the frozen state, best fit values of the parameters $a$ and $c$ show a large distribution of values, but remain largely insensitive to the quench rate of the magnet, shown in Fig. \ref{fig:params}.

\subsection{Quench rate dependence of $\chi^{\prime}_D$}
\label{sec:rqanalysis}

During our experiments, the magnetic field was decreased linearly at a set quench rate from 6~T down to 0~T and from 4~T down to 0~T, as well as between intermediate field values, as described above.  The quench rate, $r_Q$, is given by,
\begin{equation}
\begin{split}
        r_{Q} &= \frac{dB}{dt}\\
        r_{Q} &= \frac{B_{i} -B_{f}}{t_{i}-t_{f}}
\end{split}
\end{equation}
where the subscripts $i$ and $f$ refer to the initial and final parameters of the quench.  Note that $r_Q$ is defined to be positive for a decreasing field.  In order to analyze the $r_Q$ dependence of $\chi^{\prime}_D$, we considered two analytical forms.  The first is a power law, which would be potentially consistent with KZM (Eqn \ref{eq:KZM} in the main text, repeated here),

\begin{equation}
   \chi^{\prime}_{D}(r_Q) = a_p r_Q^b,
   \label{eq:powrq}
\end{equation}
where $a_p$ is a scale factor depending on the overall number of defects generated by the quenches, and $b$ is the power law exponent.  The expected values for $b$ are discussed in Appendix \ref{sec:kzm}.  

  A second analytic form accounts for the effects of domain coarsening during the quench.   This will occur for any open system where coarsening occurs on the same approximate timescale as the quench time.  We refer to the results of coarsening during the quench as a ``systematic effect'', since it does not require any additional physics such as KZM.  To derive the analytic expression we assumed that the coarsening/relaxation takes place with an identical form throughout the field range below the relevant transition, following Eqn. \ref{eqn:log_relax} (the log relaxation form).  We find that this systematic effect would lead to the following $\chi^{\prime}_D(r_Q) = \langle f \rangle - \langle l \rangle$ form at zero field (Eqn. \ref{eq:systematic} presented in the main text, repeated here):

\begin{equation}
   \chi^{\prime}_{D}(r_Q) = a\ln{\left[\frac{B_c r_Q}{t_m + B_c r_{Q}} \right]},
   \label{eq:logrq}
\end{equation}

where $t_m$ is the measurement time between the first few sections of the measured zero-field relaxation (over which $\chi^{\prime}$ is averaged to produce $\langle f \rangle$) and the last few seconds (over which $\chi^{\prime}$ is averaged to produce $\langle l \rangle$) $t_m$ was 300~s (5~min) for all of our data presented here.  Meanwhile, $a$ is in principle the same scale factor as in Eqn \ref{eqn:log_relax} (though we let it be a free parameter in our analysis and compared to the results of fitting Eqn \ref{eqn:log_relax} afterwards), and $B_c$ is a fitting parameter describing the critical field at which relaxation starts.

\begin{figure*}
    \centering
    \includegraphics{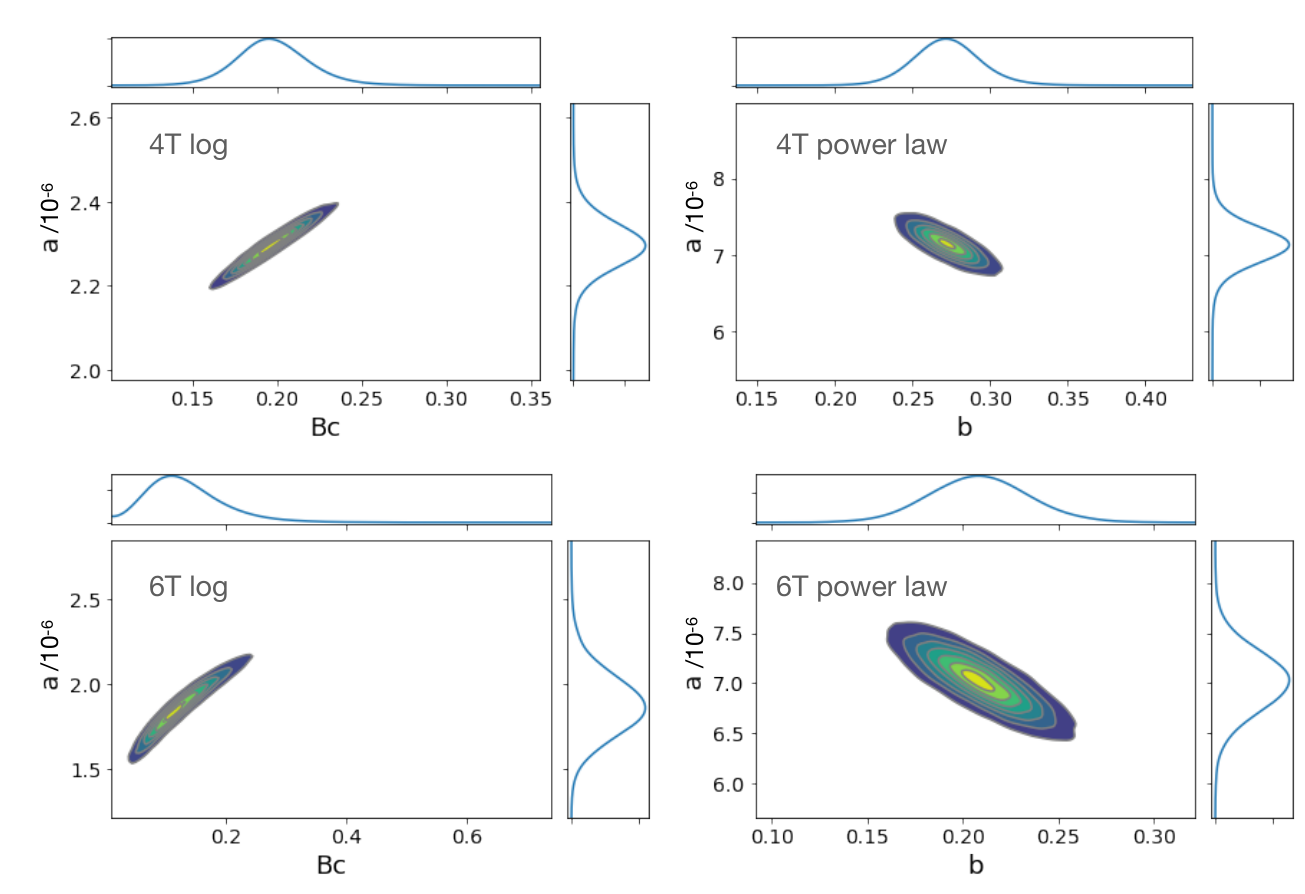}
    \caption{Joint distributions of the parameters for the two models fit to the $\chi^{\prime}_D(r_Q)$ data.}
    \label{fig:paramcorrelations}
\end{figure*}

We performed a Bayesian non-linear regression analysis of one set of 4~T quench data and one set of 6~T data.  The analysis was performed using the Markov Chain Monte Carlo package \texttt{PyMC3} \cite{pymc3} for Python.  The iPython notebook used to complete the analysis is available in the supplemental material.  We chose to use a Bayesian approach because we encountered great difficulty in reliably fitting Eqns. \ref{eq:powrq} and \ref{eq:logrq} using a traditional least squares (maximum likelihood estimation) approach, and suspected it was due to highly correlated parameters.  The advantage of Bayesian regression is that it gives access to the joint probability distribution of the model parameters, which can be inspected directly to understand the nature of the correlations (these are shown in Fig. \ref{fig:paramcorrelations} for the analyses discussed here).  Furthermore, Bayesian analysis allows for a more complete understanding of the uncertainties of the fitted models.  These are indicated by the 95\% credible intervals of the mean of the posterior distributions, shown as shaded colored areas in Fig. \ref{fig:powerlaw} panels c) and d).    For our prior distributions, we used \textit{uniform} distributions for most parameters, with limits as indicated in Table \ref{tab:priors}, except for the random error parameter (which accounts for the standard deviation of the data points away from the mean value predicted by the model), for which we used half-normal distribution.

\begin{table}[!b]
\begin{tabular}{|l|l|l|}
\hline
                       & \multicolumn{1}{c|}{4~T}                           & \multicolumn{1}{c|}{6~T}                           \\ \hline
\multirow{3}{*}{log}   & $a$: Unif(0, $5\times10^{-6}$)                    & $a$: Unif(0, $5\times10^{-6}$)                    \\
                       & $B_c$: Unif(0, 3)                                 & $B_c$: Unif(0, 3)                                 \\
                       & $\epsilon$: HalfNormal($\sigma = 1\times10^{-5}$) & $\epsilon$: HalfNormal($\sigma = 1\times10^{-5}$) \\ \hline
\multirow{3}{*}{power} & $a$: Unif(0, $40\times10^{-6}$)                   & $a$: Unif(0, $40\times10^{-6}$)                   \\
                       & $b$: Unif(0, 2)                                   & $b$: Unif(0, 2)                                   \\
                       & $\epsilon$: HalfNormal($\sigma = 5$)              & $\epsilon$: HalfNormal($\sigma = 5$)              \\ \hline
\end{tabular}
\caption{Prior distributions for the model parameters used in the Bayesian regression. $\epsilon$ is the standard deviation associated with the normal distribution of the data away from the mean (best-fit line).  We chose a half-normal distribution centered at 0 since the standard deviation cannot be negative.}
\label{tab:priors}
\end{table}

\bibliography{CoNbO_bib.bib}

\end{document}